# Engineering chromium related single photon emitters in single crystal diamond

**I Aharonovich, S Castelletto, B C Johnson, J C McCallum and S Prawer**

School of Physics, University of Melbourne, Parkville, Victoria 3010, Australia

Email address: i.aharonovich@pgrad.unimelb.edu.au

**Abstract.** Color centers in diamond as single photon emitters, are leading candidates for future quantum devices due to their room temperature operation and photostability. The recently discovered chromium related centers are particularly attractive since they possess narrow bandwidth emission and a very short lifetime. In this paper we investigate the fabrication methodologies to engineer these centers in monolithic diamond. We show that the emitters can be successfully fabricated by ion implantation of chromium in conjunction with oxygen or sulfur. Furthermore, our results indicate that the background nitrogen concentration is an important parameter, which governs the probability of success to generate these centers.

Developing novel solid state systems which can generate single photons on demand at room temperature is a recognized goal in the quantum information science community [1, 2]. Impurities in diamond offer a unique advantage over other systems as some of them are optically active and can be employed as true single photon sources [3]. The nitrogen vacancy (NV) center for example[4] has been the subject of intense research[4,5], particularly due to its optical spin readout capabilities and its potential use as a high sensitivity magnetic sensor[6, 7]. However, the optical properties of this center are limited by a strong phonon coupling, resulting in a broad emission (~100 nm) of which only 4% is concentrated in the zero phonon line (ZPL). This is a significant drawback for many quantum optical applications including quantum key distribution, quantum metrology and optical quantum computation.

Alternative centers with narrower emission lines which are also more suitable for microcavity integration are required. Nickel and silicon related emitters have been studied as plausible candidates since some of the centers show narrow emission lines in the near infra-red (NIR)[8-11]. However, nickel centers such as NE8 are difficult to fabricate due to the requirement of four nitrogen atoms to form a complex with the Ni impurity. Until recently the silicon vacancy (SiV) has not been considered as a good candidate for single defect devices due to a very high non radiative decay term, limiting its quantum efficiency to only 5%. Surprisingly, however, the recent fabrication of SiV in chemical vapor deposition (CVD) grown nanodiamonds formed on an iridium substrate demonstrated an outstanding and unexpected brightness and spectral properties[12]. These improvements are likely to spur substantial research into the photophysics of the SiV center, particularly, fabricating centers with similar brightness in monolithic diamond by ion implantation.

Recently a new class of single photon emitters associated with chromium impurities in diamond[13-15] was discovered. These centers show bright fluorescence in the NIR and possess a short radiative lifetime. Moreover the linear transition dipole both in absorption and in emission[16] is considered ideal for 3D orientation imaging and tracking with various microscopy methods[17-19].

The chromium emitters were originally discovered during diamond growth on a sapphire substrate. The Cr atoms which are present in the sapphire substrate, were incorporated into the growing crystal through gas phase diffusion[9]. However, to achieve the best intrinsic photo-physical properties of optical centers, such as reduced inhomogeneous broadening and Fourier transform limited emission, emitters should be fabricated into single crystal diamond[20]. Therefore, methods to create optical centers with high efficiency in bulk material are of paramount importance for quantum optical devices[21, 22]. Furthermore, for scalable quantum device architectures and for integration with other optical structures, accurate positioning and a high formation probability are required. These conditions have not yet been fulfilled for the particular case of chromium centers. Above all, the role of other impurities in the formation of the center remains poorly understood. A formation study of this center by conventional ion implantation techniques is also required in order to assess the possible deterministic fabrication pathways to engineer the centers in nanodiamonds sized below 50 nm. This would have immediate application in bio-sensing.

In this paper we investigate the role of co-dopants such as nitrogen and oxygen, on the formation of chromium centers in diamond by ion implantation techniques. We perform a range of co-implantations of various impurities together with Cr ions, to determine their effect on the formation of the centers at the single emitter level having superior optical properties as described in [15]. We also consider the target material and show that the background nitrogen level present natively inside the diamond is a crucial parameter. Our results aim to provide a sound basis for further research into fabrication of the chromium emitters and gather important information regarding their atomistic structure. We envisage that the results will encourage further experiments to improve the formation yield of the centers up to the desired level necessary to implement deterministic devices based on Cr single emitters in diamond[21].

The ion implantation of chromium, oxygen, silicon, boron, nitrogen and sulfur was performed with a National Electrostatics Corporation 150 kV low energy ion implanter at the Australian National University. During implantation the single crystal diamond samples (3×3×0.5 mm$^3$) were affixed to a temperature controllable nickel block with clamps. All implantations were performed at room temperature in ultra high vacuum conditions (~10$^{-7}$ Torr). The implantation ion energies for chromium and the co-implanted species were all chosen to result in a projected range of 25 nm and are listed in Table 1. This allows efficient detection of the light emitted from the centers. Figure 1 shows the concentration profiles of the implanted ions as determined with the Monte Carlo simulation code Stopping and Range of ions in matter (SRIM)[23].

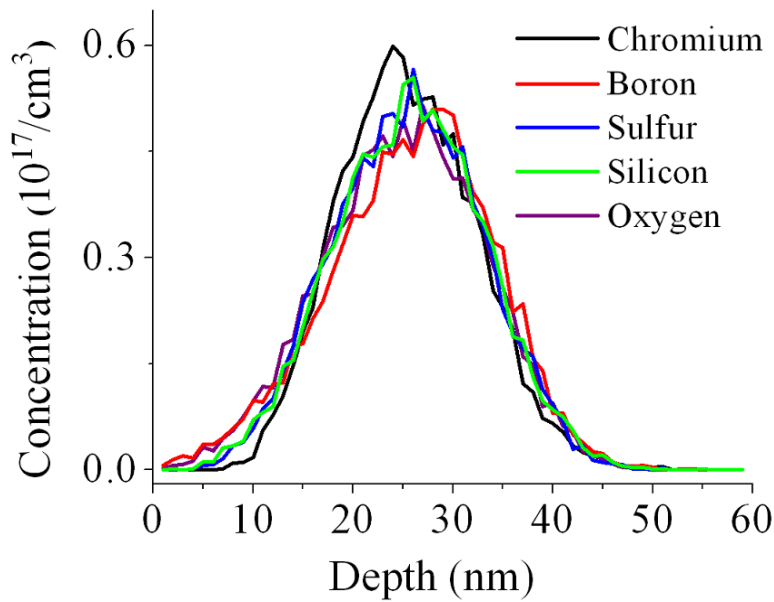

Figure 1. SRIM simulation of the concentration profiles of the implanted ions.

Table 1. Implanted elements and the corresponding implantation energies and projected ranges.

| Element | Implantation energy [keV] | Projected Range [nm] |
|---|---|---|
| Chromium | 50 | 25±7 |
| Oxygen | 19.5 | 25±8 |
| Silicon | 34 | 25±7 |
| Boron | 13 | 25±8 |
| Sulfur | 38 | 25±7 |
| Nitrogen | 18 | 25±8 |

After implantation, the samples were annealed at 1000°C in a forming gas ambient (95% Ar – 5% $H_2$) for 2 hours. This is a typical procedure to induce vacancy diffusion in diamond and to repair damage caused by the implantation. Note that the annealing step applied after the implantation is not sufficient to cause any diffusion of the implanted atoms (Cr, Si, B, O, N) in the diamond lattice[24-27].

The samples were then optically characterized to identify chromium related single emitters. This was performed using a home built confocal microscope with a Hanbury Brown and Twiss (HBT) setup to gain information regarding the photon statistics of the centers. A fiber coupled continuous wave diode laser emitting at 682 nm was used for excitation. The diamond sample was mounted on a piezo XYZ stage with 0.2 nm resolution, allowing 100×100 $\mu m^2$ scans. The emitted light was collected using a high numerical aperture objective (NA=0.9) and coupled into a 62.5 μm core multimode fiber, which acts as an aperture. A 50:50 fiber-coupled beam splitter guided the photons to two single photon counting detectors (APDs) and their outputs were sent to the start and stop inputs of a time correlator card. All the measurements were performed at room temperature. Spectroscopy has been performed to determine the typical Cr emission lines associated to a particular bright spot on the confocal map.

We first investigated co-implantation of chromium with oxygen into type IIA CVD ([N] < 1 ppm, [B] < 0.05 ppm) samples. This was prompted by the fact that chromium in sapphire needs to be in the 3+ oxidation state to be optically active[28]. Similarly, research on incorporation of Er into silicon found that co-implantation of oxygen increased the proportion of Er that was present in the optically-active 3+ state [29, 30, 31]. This approach has not been explored for chromium in silicon, probably because the luminescence would most likely not be in a useful energy range. Instead research on chromium in silicon has been explored in the context of transition metal dopant interactions, for example, formation of CrB pairs[25, 26]. The experiments based on co-implantation of chromium and oxygen or selected other species (Si, S, B) are summarized in Table 2. Figure 2 shows the main experimental results graphically.

Table 2. Summary of chromium related ion implantations into type IIA CVD ([N] < 1 ppm, [B] < 0.05 ppm) diamond.

| Diamond material | Fluences (ions/cm$^2$) | Sample # | Yield (average area density of Single Cr/implanted ion)% |
|---|---|---|---|
| Type IIA [N] < 1 ppm [B] < 0.05 ppm | 1×10$^{11}$ Cr+1×10$^{11}$ Si | (1) | 10$^{-5}$ |
| | 1×10$^{13}$ Cr+1×10$^{13}$ Si | (2) | 10$^{-7}$ |
| | 1×10$^{11}$ Cr | (3) | 10$^{-5}$ |
| | 1×10$^{13}$ Cr | (4) | 10$^{-7}$ |
| | 1×10$^{10}$ Cr+1.5×10$^{10}$ O | (5) | 10$^{-3}$ |
| | 2×10$^{10}$ Cr+3×10$^{10}$ O | (6) | 5×10$^{-4}$ |
| | 1×10$^{11}$ Cr+1.5×10$^{11}$ O | (7) | 10$^{-4}$ |
| | 1×10$^{11}$ Cr+5×10$^{11}$ O | (8) | 10$^{-4}$ |
| | 1×10$^{11}$ Cr+2×10$^{12}$ O | (9) | 10$^{-4}$ |
| | 1×10$^{12}$ Cr+1.5×10$^{12}$ O | (10) | 10$^{-5}$ |
| | 1×10$^{13}$ Cr+1.5×10$^{13}$ O | (11) | 10$^{-7}$ |
| | 1×10$^{11}$ Cr+3×10$^{11}$ S | (12) | 10$^{-4}$ |
| | 1×10$^{11}$ Cr+1×10$^{11}$ B | (13) | 10$^{-5}$ |
| | 1×10$^{11}$ Cr+1×10$^{12}$ B | (14) | 10$^{-5}$ |

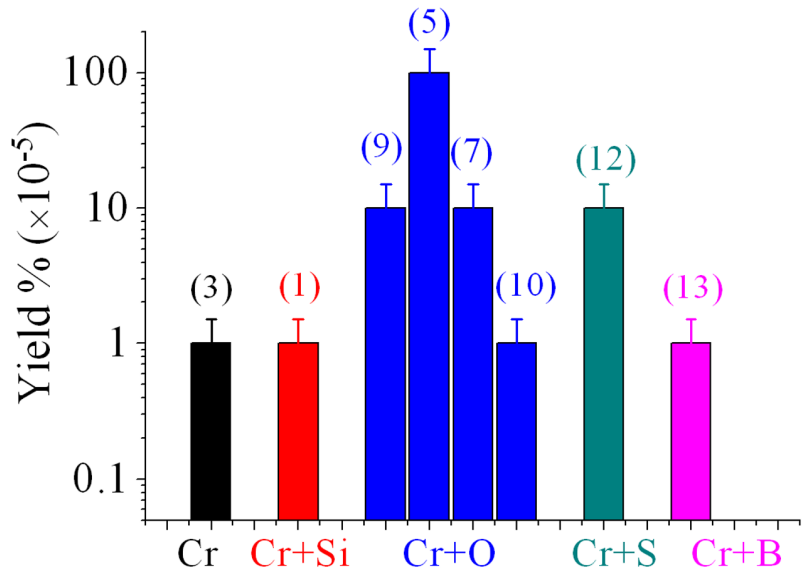

Figure 2. Various implantation routes employed to investigate the formation of Cr related centers in type IIA CVD ([N] < 1 ppm, [B] < 0.05 ppm) diamond. The number in the brackets denotes the sample number, as shown in table 1. The yield is the average area density of single Cr centers over the implanted ion fluences.

Figure 3(a,b) shows typical confocal maps recorded from samples 3 and 7, respectively. It is clearly seen that more bright centers can be found on sample 7 which was co-implanted with oxygen rather than sample 3, which was implanted with chromium only.

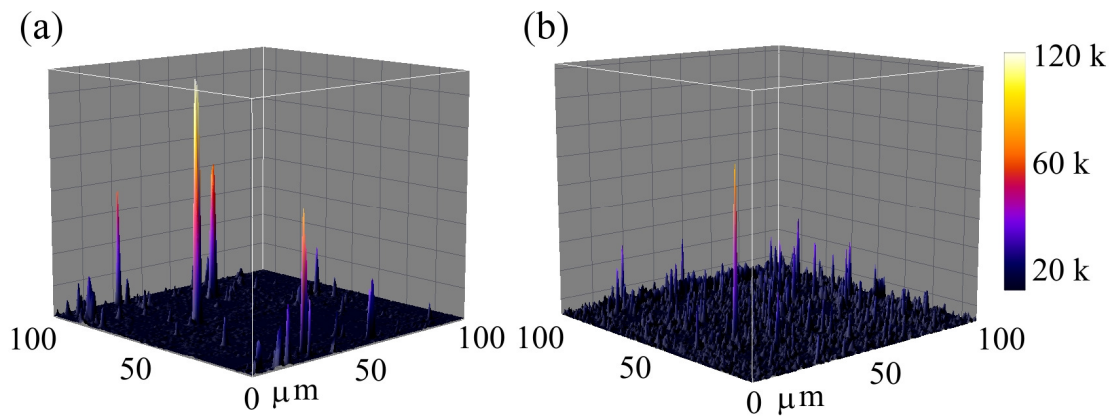

Figure 3. Confocal map recorded from a diamond sample implanted with (a) chromium and oxygen (sample 7) and (b) chromium only (sample 3). More single emitters are clearly observed at sample 7, which had chromium co-implanted with oxygen.

Figure 4a shows a typical spectrum of the chromium related center. Inset, is the anti-bunching measurement recorded from the same emitter employing the HBT setup demonstrating that the center is a single photon emitter. Narrow bandwidth lines with FWHM ~4 nm were routinely observed from the chromium implanted samples. The histogram of the ZPLs is shown in figure 4b. Similar ZPLs were observed in samples (1-14).

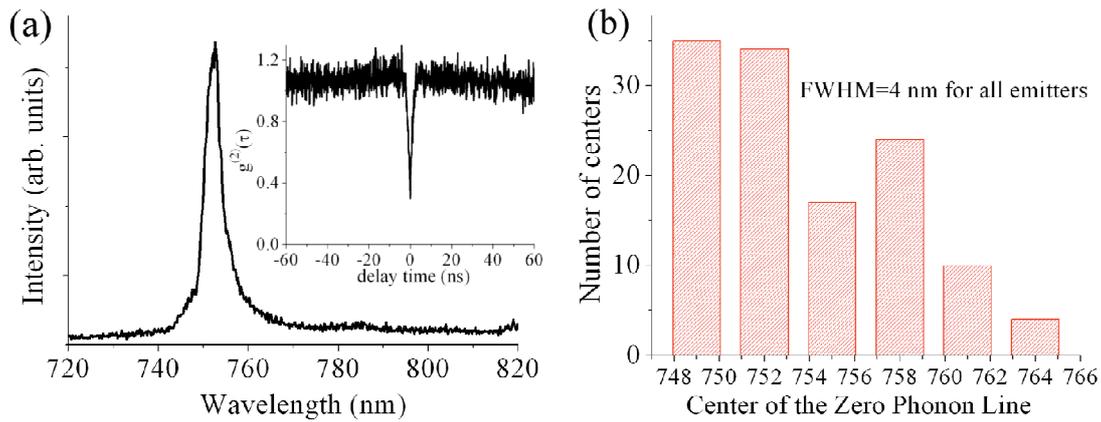

Figure 4 (a) PL spectrum of a chromium related emitter. Inset is the $g^{(2)}(\tau)$ function recorded from the same emitter demonstrating that the addressed center is a single photon emitter. (b) Histogram of the ZPLs of chromium related emitters found in different samples.

Implantation of $1\times10^{11}$ Cr/cm$^2$ into type IIA CVD diamond (Sample 3) yields on average 1 optically-active Cr centre per 100×100 μm$^2$ scan. This indicates a rather low conversion probability of implanted Cr into optically active defects. Increasing the Cr fluence by two orders of magnitude decreases the yield by two orders of magnitude (Sample 4). This may indicate that residual implantation damage plays a role in limiting the yield of the centers.[27] The increased damage at higher fluences could further reduce the impurity diffusivities [26] and/or reduce the probability of impurities attaining the required charge state. Co-implantation of Cr and O increases the yield by an order of magnitude or more (Samples 5-9), but again, an increased fluence results in a reduced yield. The highest yield was obtained for $1\times10^{10}$ Cr+$1.5\times10^{10}$ O ions/cm$^2$ (Sample 5). It is important to note that the chromium centers were repeatedly fabricated in type IIA CVD ([N] < 1 ppm, [B] < 0.05 ppm) diamond but could not be formed in other types of diamond substrates as discussed further below.

Both oxygen and sulfur increase the yield of optically-active Cr centers (samples 5-12). The ratio of chromium related single emitters observed in the co-implantation of chromium and oxygen/sulfur compared to other implantations was as high as 10:1. Since sulfur and oxygen are located in the same column in the periodic table, they are expected to behave similarly when introduced into the diamond lattice. Sulfur defects in diamond have been investigated as potential donors to achieve n-type material. It is assumed that sulfur occupies a substitutional lattice site in diamond[27]. Theoretical work also predicts that oxygen will occupy a substitutional lattice site[28].

The general low formation probability of all implantation schedules studied is consistent with an expectation that the Cr and oxygen (or sulfur) atoms must be in very close proximity following the implantation process for the appropriate optically active charge state to form during the subsequent annealing, especially since the diffusivities are expected to be very low in the annealing regime used.

As noted earlier, the role of oxygen and sulfur in the formation of optically-active Cr centers in diamond may be similar to the role O plays in Er doping of silicon. In silicon, under the processing conditions used, the diffusivity of oxygen is high enough that oxygen present in the silicon bulk can combine with a proportion of the implanted Er to form the optically-active centre. Co-implantation enhances the $Er^{3+}$ emission. EXAFS measurements on Er- and O-doped Si have shown that Er is coordinated by four to six O atoms[29]. Direct bonding is therefore important in forming the optically-active center in this system.

Co-implantation of silicon and chromium was performed to test the damage effect associated with the implantation. Furthermore, silicon should not modify the charge state since it has a similar electronic configuration as carbon. Instead, the Si will only produce damage and introduce more vacancies. This implantation procedure followed by the standard annealing treatment did not enhance the number of Cr centers (samples 1-2). The yield was not modified appreciably from the samples implanted only with Cr. These results confirm that the role of the oxygen/sulfur is related to a charge transfer rather than assisting in the generation of vacancies.

To investigate the effect of p-doping, chromium and boron were co-implanted (samples 13-14). Also CrB pairs are speculated to be responsible for some optically active centers in silicon when Cr is present, therefore, there is a scientific interest to investigate this pair in diamond[25, 26]. The results of these implants showed that only a limited number of chromium emitters were formed, similar to the chromium only or chromium plus silicon co-implantation schedules. The charge state configuration could be aided by as-grown impurities (e.g. from nitrogen). However, as was shown experimentally, the probability of forming the chromium emitters is significantly reduced.

For the samples implanted with $1\times10^{11}$ Cr/cm$^2$ no significant oxygen fluence dependence was observed for the Cr:O ratio 1:1.5 to 1:20 (Samples 7-9) considered. The optical properties of the centers were likewise unaffected by the greater oxygen concentration.

In the second part of our work, we investigated the target diamond material dependence on Cr center formation. Three types of diamond were compared: CVD single crystal type IIA diamond ([N] < 1 ppm, [B] < 0.05 ppm), ultra pure CVD single crystal diamond ([N] < 5 ppb, [B] < 1 ppb), both purchased from Element Six and a type IB sample ([N] < 100 ppm, [B] < 0.1 ppm) produced by Sumitomo. This last crystal was cut and polished from a large single crystal which was synthesized under high pressure high temperature (HPHT) conditions. Table 2 summarizes the performed experiments. The annealing sequence was the same as described above.

Table 3. Various implantation routes employed to investigate the formation of Cr related centers in an ultra pure CVD diamond ([N] < 5 ppb, [B] < 1 ppb), and in a Type Ib diamond ([N] <100 ppm, [B] < 0.1 ppm) grown by HPHT method.

| Diamond type | Implantation details | Sample # | Comments |
|---|---|---|---|
| Ultra pure [N] < 5 ppb, [B] < 1 ppb | $1\times10^{11}$ Cr+$1.5\times10^{11}$ O | 15 | No single centers or narrow PL lines were found at all |
| | $2\times10^{11}$ Cr+$3\times10^{11}$ O | 16 | |
| | $2\times10^{10}$ Cr | 17 | |
| | $2\times10^{10}$ Cr+$2\times10^{10}$ B | 18 | |
| | $1\times10^{11}$ Cr+$1.5\times10^{11}$ O + $1\times10^{9}$ N | 19 | |
| | $1\times10^{11}$ Cr+$1.5\times10^{11}$ O + $1\times10^{11}$ N | 20 | |
| Type Ib [N] <100 ppm [B] < 0.1 ppm | $1\times10^{11}$ Cr+$1\times10^{11}$ Si | 21 | No single centers or narrow PL lines were found at all. |
| | $1\times10^{13}$ Cr+$1\times10^{13}$ Si | 22 | |
| | $1\times10^{11}$ Cr | 23 | |
| | $1\times10^{13}$ Cr | 24 | |
| | $1\times10^{11}$ Cr+$1.5\times10^{11}$ O | 23 | |
| | $1\times10^{13}$ Cr+$1.5\times10^{13}$ O | 25 | |

The results of this section are intriguing. When the implantation sequence which generated single emitters in type IIA diamond (e.g. $1\times10^{11}$ Cr+$1.5\times10^{11}$ O) was applied to an ultra pure or a type Ib diamond, single chromium related centers were never observed (samples 15-18, 21-25). Figure 5 shows a PL spectrum recorded from the ultra pure diamond sample (sample 15). A broad emission is observed, similar to the Cr related cathodoluminescence map reported in Ref [32]. Neither narrow bandwidth emission nor single photon characteristics were observed from this sample.

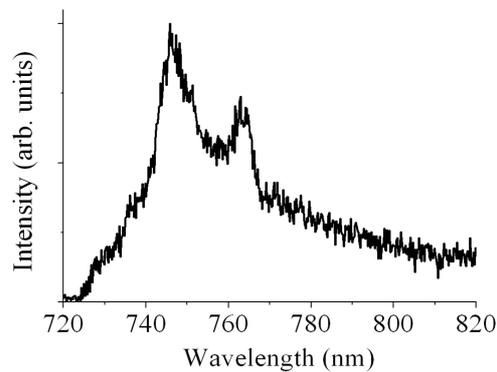

Figure 5. PL spectrum recorded from chromium and oxygen implanted ultra pure diamond.

The main difference between these diamonds is the amount of nitrogen. While in an ultra pure sample, the level of nitrogen is extremely low, less than ppb, in type Ib the concentration of nitrogen is extremely high (~100 ppm). It is therefore likely that nitrogen plays a crucial role in the formation of the centers. The influence of nitrogen is likely to dominate the charge state of the center. If its concentration is too high or too low, the chromium emitters may not be formed because the correct charge state cannot be achieved. Theoretical work by Gali et al.[33] indicates that the charge state of substitutional oxygen depends very sensitively on the Fermi level so this may also play a role.

To investigate this hypothesis further, we performed two additional experiments: (1) We co-implanted nitrogen with chromium and oxygen (samples 19-20). The nitrogen fluence was chosen to match the ~1 ppm of native nitrogen available in type IIA diamonds. (2) Sample 20 was annealed and re-implanted with chromium. This was done to achieve stabilization of the implanted nitrogen so that the nitrogen atoms can occupy the most stable location in the diamond lattice. However, even in this case no single centers or narrow PL lines were observed. This stresses the importance of having a particular initial nitrogen concentration in the target diamond sample. Note that only a small fraction of the implanted nitrogen atoms occupy substitutional lattice sites and form substitutional nitrogen defects. The rest will bond to a vacancy forming NV centers. Moreover nitrogen-interstitial complexes can be produced in significant concentrations due to N implantation, and can act as electron traps and are stable even at temperatures where the nitrogen-vacancy centre anneals out [34]. Full activation of the nitrogen will require higher annealing temperatures currently not available in our laboratory.

In the last part of our experiments we performed high energy ion implantation of 9 MeV chromium and 5.1 MeV oxygen to fluences of $1\times10^{11}$ and $1.5\times10^{11}$ ion/cm$^2$ respectively on a NEC 1.7 MV tandem accelerator. After the same annealing sequence, the samples were investigated optically. The results of the deep implantation were similar to the shallow one, with an average yield of 5-10 centers in a 100x100 μm$^2$ area. This observation strongly suggests that vacancies are not the limiting factor in the formation of the center, since more than 5000 vacancies/ion are formed during the high energy implantation. This is much greater than that produced by the low energy implanted ions. On the other hand, it was shown recently that NV formation dramatically increases when the nitrogen is implanted deep into the diamond. The higher the implantation energy, the more vacancies are created around the substitutional nitrogen atom, giving rise to a higher yield of NV centers[35]. Therefore, even if vacancies are not forming the chromium center, a more radical condition (such as a specific charge state or complex with additional impurity) must be met to form the centers.

To summarize, by employing ion implantation and confocal microscopy, we investigated the fabrication methodologies of chromium related centers and the influence of co-implanted impurities. Our results clearly show that the centers can be routinely fabricated in CVD type IIA single crystal diamond. The best recipe for producing chromium single photon emitters is implanting chromium and either oxygen or sulphur to a fluence of ~$1\times10^{10}$ ions/cm$^2$ followed by a two hour, 1000$^o$C anneal in a forming gas ambient. To continue the progress in understanding the physical properties of the new family of chromium defects, further experimental and theoretical work is required[36]. Once solid theoretical support is provided, it may be possible to increase the fabrication yield of the centers and fundamental research mapping out the spin properties and atomic composition of the defects can be launched.

**Acknowledgments**

The Department of Electronic Materials Engineering at the Australian National University is acknowledged for providing access to ion implanting facilities. This work was supported by The International Science Linkages Program of the Australian Department of Innovation, Industry,